# Evaluating Privacy Questions From Stack Overflow: Can ChatGPT Compete?

Zack Delile, Sean Radel, Joe Godinez, Garrett Engstrom, Theo Brucker, Kenzie Young, Sepideh Ghanavati
*School of Computing and Information Science, University of Maine, Orono, ME, USA*
{zachary.delile, sean.radel, joseph.godinez, garrett.engstrom, theodore.brucker, mackenzie.k.young, sepideh.ghanavati}@maine.edu

*Abstract*—Stack Overflow and other similar forums are used commonly by developers to seek answers for their software development as well as privacy-related concerns. Recently, ChatGPT has been used as an alternative to generate code or produce responses to developers' questions. In this paper, we aim to understand developers' privacy challenges by evaluating the types of privacy-related questions asked on Stack Overflow. We then conduct a comparative analysis between the accepted responses given by Stack Overflow users and the responses produced by ChatGPT for those extracted questions to identify if ChatGPT could serve as a viable alternative. Our results show that most privacy-related questions are related to choice/consent, aggregation, and identification. Furthermore, our findings illustrate that ChatGPT generates similarly correct responses for about 56% of questions while for the rest of the responses, the answers from Stack Overflow are slightly more accurate than ChatGPT.

## I. INTRODUCTION

Recent studies examine requirements analysts' and developers' privacy behaviors [1]–[6] and show that analysts and developers generally see privacy as a burden and afterthought and are not familiar with basic privacy concepts [7]–[9]. However, with the increase in the number of privacy regulations around the world [10], [11], analysts and developers, especially in small companies with little access to privacy and legal experts, seek to address their privacy-related questions through forums such as Stack Overflow or Reddit [12], [13]. More recently generative AI-based tools such as ChatGPT [14], [15] have gained more popularity among developers, as well.

Previous research evaluates privacy-related questions on Stack Overflow and Reddit [16]–[19] to understand analysts' and developers' challenges and the trends towards privacy-related questions. These works show that most questions revolve around privacy regulations and a few of the privacy by design strategies [20]. These efforts provide insights regarding privacy trends and questions; however, their scopes are still very limited and focus only on keywords to identify if the topics of the questions are privacy-related or not.

ChatGPT[1], a tool introduced by OpenAI [21], is based on a GPT architecture [22] and has received much attention in the software engineering domain, among others [14], [15]. Through prompt engineering, ChatGPT can produce responses to prompts (i.e., questions), have a conversation, and be used in identifying requirements ambiguities [23], code generation and completion tasks [24], bug fixing [25] and software testing [14]. With this trend, ChatGPT may replace or complement these forums for software- or privacy-related questions.

In this paper, we seek to assess whether ChatGPT could serve as a viable alternative to Stack Overflow in answering privacy-related questions and generating similarly correct results. To achieve this goal, we extend the current efforts on evaluating developers' forums through a novel methodology, inspired by approaches in Natural Language Inference (NLI) [26], by extracting and classifying the types of Stack Overflow privacy-related questions based on a set of premises and hypotheses. We then conduct a comparative analysis with the responses received from ChatGPT.

For this, we randomly select 270 questions and answers (QAs) from 932 questions extracted between 2016 and 2023 from Stack Overflow. Next, we develop an annotation strategy to classify the types of QAs based on two well-known privacy taxonomies [27]. After four rounds of annotations with three groups of two annotators (i.e., six in total), and two rounds of discussions to resolve the differences, we create a dataset of 92 pairs of multi-labeled privacy-related QAs. The results of our annotations show that the questions contain concepts of choice/consent, aggregation, identification, disclosure, and onward transfer more often than the other concepts. This result indicates the importance of informing users, confidentiality, and sharing with third parties, for developers. In addition, the collection limitation concept is always linked to choice/consent whereas disclosure is only labeled together with choice/consent in 63% of cases, showing that developers understand the need for getting consent before data collection or disclosure.

We then follow prompt engineering [28] and generate answers for Stack Overflow privacy-related questions in ChatGPT and develop a study to identify whether the solutions from ChatGPT match with the accepted answers or the highest upvoted answers from Stack Overflow. Our results show that when given the same prompt, ChatGPT responses match with Stack Overflow answers in 56.1% of cases. When it does not match, SO is slightly more accurate than ChatGPT. This result indicates that ChatGPT may in the future be used as an alternative to SO and other forums. However, since both of these tools show an accuracy of less than 75%, developers need to use the forums and ChatGPT with caution to ensure compliance with regulations and protect users' privacy.

---
[1] https://openai.com/blog/chatgpt

## II. RELATED WORK AND BACKGROUND

Stack Overflow (SO) [12], its variants and Reddit [13] are forums for developers to ask their software-related questions. With the introduction of privacy regulations [10], [11] requirements analysts and developers, especially in smaller companies, turn to these forums to ask their privacy-related questions. A recent analysis of 727 publications since 2010 [29] shows that Reddit has become a substantially more popular medium in various disciplines, including computer science and software development processes.

To understand how developers use these forums and the types of privacy-related questions they ask, some work focus on examining Stack Overflow and Reddit [16]–[19], [29], [30]. Analysis of subreddits of the top 10 software applications [31] shows that ~54% of the posts contain useful information for requirements elicitation processes such as bug reports or feature requests. Li et al. [18] analyzed 207 discussions on r/androiddev subreddit and found that even though the discussions regarding data protection and privacy are not as prevalent, there is an emerging trend toward these types of questions motivated by compliance with the new regulations.

Tahaei et al. [16], [17], [32] conducted multiple studies regarding privacy-related questions [16], the types of privacy advice given by developers on SO [17], as well as questions regarding permissions specific to health applications [32]. In one study, they evaluated 315 privacy questions on SO and identified that Google and Apple Privacy Labels positively impacted the number of privacy questions asked [16], especially regarding informing the users. In [32], they evaluated 269 privacy-related questions for health applications and identified that the majority of the concerns are related to permissions scope, authentication, and third-party library usage. Lastly, they examined 148 pieces of advice extracted from SO against Hoepman's privacy by design approach [20] and identified that they mainly focus on strategies related to confidentiality or compliance such as 'inform,' 'hide,' 'control,' and 'minimize.'

The introduction of ChatGPT [21], supported by GPT-3.5 architecture [22], has gained significant interest among the software engineering community in achieving tasks such as code completion and generation [15] and bug fixing [25]. Many studies focus on evaluating the performance of ChatGPT on requirements and software engineering activities [15], [25], [33] but little work focus on expert-intensive tasks such as *privacy*. As a programming tool, the model can solve most easy and medium-level problems and potentially surpass the ability of a novice-level programmer [14]. ChatGPT, with the ability to learn from the user querying the model, can increase the success rate of bug fixing to 77.5%, which outperforms other automated program repair methods [25]. While ChatGPT can successfully generate code to solve a problem, it struggles to solve new or unseen problems [15].

Research shows that there is a lack of empirical approaches to comparing software engineers' and requirements analysts' ability to solve technical problems versus machine-learning tools [34]. In this paper, we aim to tackle this gap by evaluating ChatGPT's ability to generate privacy-related responses in comparison to users' answers on Stack Overflow and providing insights about its performance on such expert-intensive tasks regarding privacy concerns.

## III. METHODOLOGY

In this section, we first describe our methodology for creating the dataset of privacy-related QAs from Stack Overflow (SO), and then we discuss our approach for comparing the quality of the answers given by SO users versus ChatGPT.

### A. Data Collection and Annotation Process

To gather the relevant data, we first extracted 932 questions from SO posted between 2016 and 2023 and randomly selected 270 QA pairs from the list for our annotation and analysis. We divided these 270 entries into three files of 90 entries and assigned each file to two annotators who have privacy expertise. We developed an annotation guideline and a template (see Figure 2) and then performed four rounds of the annotation process and two rounds of discussions to evaluate whether an entry is privacy related or not. We leverage an approach from Natural Language Inference (NLI) [27], [35], [36] to identify if an entry is privacy-related or not. Harkous et al. [27] developed a set of hypotheses based on two well-known taxonomies [37], [38] to help identify if an app review entails privacy concepts. We use and leverage their list for our task. Figure 1 shows an excerpt of hypotheses from both taxonomies. The complete list is presented in [27]. In NLI, we can define three types of relationships (i.e., as *entailment*, *neutral* and *contradiction* [39]) between each premise (here, the SO questions) and a hypothesis. In our task, if a statement is an *entailment*, we consider them privacy-related; otherwise, we leave them out.

| Privacy Concepts | Hypothesis |
|---|---|
| **Concepts from Solove's Taxonomy** | |
| Surveillance | The user is facing a data surveillance issue. |
| Aggregation | Personal user information is collected from other sources. |
| Identification | A data anonymity topic is discussed. |
| Secondary Use | The user is concerned about the purposes of personal data access. |
| Exclusion | The user wants to correct their personal information. |
| Breach of Confidentiality | A breach of data confidentiality is discussed. |
| Disclosure | Personal data disclosure is discussed. |
| **Concepts from Wang and Kobsa's Taxonomy** | |
| Notice/Awareness | Opting out from personal data collection is discussed. |
| Data Minimization | More access than needed is required. |
| Purpose Specification | The reason for data access is not provided. |
| Collection Limitation | Too much personal data is collected. |
| Use Limitation | The data is being used for unexpected purposes. |
| Onward Transfer | Data sharing with third parties is discussed. |
| Choice/Consent | User choice for personal data collection is discussed. User did not allow access to their personal data. |

Fig. 1. An Excerpt of the Hypotheses List [27]

**Annotation – Round_1 & Discussion_1:** In Round_1, each group reviewed the first 10 questions in their file. They checked the entries against the list of hypotheses and labeled a question as *privacy* if it had an *entailment* relationship.

After the initial annotation, we conducted a discussion session with all annotators to identify the challenges in the process and to improve our methodology. We created a guideline and a template (Figure 2) to help facilitate the annotation, comparison, and analysis in the later steps. We also created a questionnaire for feedback about the annotation process after Rounds_2 and _3 for possible improvements.

**Annotation – Round_2 to Round_4:** After creating the template and the guidelines, we conducted three additional rounds of annotation with each group for all 90 entries in each file, including the original 10 questions reviewed in Round_1. Round_2 and _3 included 20 questions per group of annotators while Round_4 included the last 50 entries.

For each question that is labeled as privacy, the annotators also determined the types of privacy concepts based on the attributes shown on the left-hand side of Tables III and IV. The annotators labeled each entry with one or more attributes leading to a multi-label dataset. For each question, the annotators were also required to save the accepted answer, the number of votes for the accepted answer, the answer with the highest number of votes, and the number of said votes.

Fig. 2. An Excerpt of the Annotation Template

In between each round of annotation, the annotators were given a questionnaire to suggest feedback regarding the methodology for potential improvement in future rounds. Ultimately, the annotation method was kept the same in all three rounds, and 270 questions were annotated. At the end of these rounds, 95 questions were labeled as *privacy* by at least one annotator while 54 of them were annotated by both. Round_2 and _3 took an average of 40 – 60 minutes, while Round_4 took an average of 90 – 120 minutes, per annotator.

**Annotation – Discussion_2:** Upon completion of the annotation rounds, we calculated the Cohen-Kappa interrater agreement[2]. We then conducted three pair-wise discussion sessions where we selected the questions that were annotated as *privacy* by at least one of the two annotators in each group. Each pair of annotators were tasked to examine their individual annotation and reach an agreement with their pair regarding the privacy-related questions. At the end of this process, our dataset included 92 questions and their answers. The categories are shown on the left-hand side of Tables III and IV.

### B. Comparative Analysis with ChatGPT

As mentioned earlier, one of our objectives is to verify whether ChatGPT could serve as an alternative for developers with expert-needed questions such as privacy. Hence, in the second part of our study, we conduct a comparative analysis between the Stack Overflow answers (SOA) and the ChatGPT-generated answer (GPT-A). We use the free version of ChatGPT, which is used on the GPT-3.5 model since it is the most used version. For the prompts, we use the questions extracted from SO *as-is* for a direct comparison. Figure 3 shows an example of a GPT-A generated by the first author.

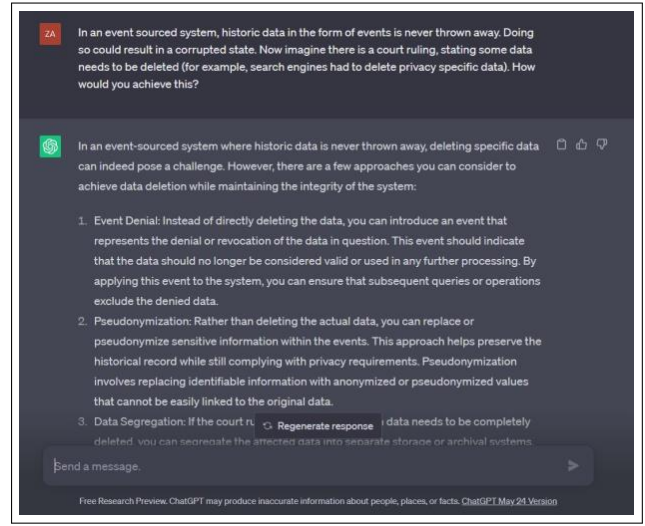

Fig. 3. An Excerpt of Generated GPT-A

During our annotation process, we collected the accepted answers and/or the answers with the highest number of upvotes. We used these two sets of answers to create the SOA dataset. If a question included an accepted answer, we added that to the SOA dataset. In cases where a question had no accepted answer, we added the answer with the highest number of upvotes to the SOA dataset. Out of our 92 questions, 10 questions had no answer; thus, the SOA dataset includes only 82 questions/answers. Next, the first author generated the GPT-A for each question by copying and pasting the question text as a prompt in ChatGPT. The first answer provided by ChatGPT was added to the GPT-A dataset.

To conduct the analysis, we split the questions among five out of the original six annotators with privacy expertise, with each question being evaluated by three annotators. We eliminated the first annotator from this task to prevent any potential biases since the annotator was responsible for generating ChatGPT responses. The annotators were given the template shown in Figure 4 where the SOA and the GPT-A datasets were presented in an anonymized format. To mitigate the bias, they did not know which answer is from SOA and which is from GPT-A. They were then asked if the two answers agreed with each other. If an annotator said yes, they were asked whether the answers were an exact match or a similar match. A similar match means that the answers are close enough but use synonyms or alternative solutions. If an annotator said no, they were asked which answer seems to be more accurate based on their expertise. We calculate Fleiss-Kappa interrater agreement[3] and report the results in Section IV.

---
[2]https://www.statisticshowto.com/cohens-kappa-statistic/

[3]https://www.statisticshowto.com/fleiss-kappa/

Fig. 4. An Excerpt of the Template for the Comparative Analysis

TABLE I
COHEN-KAPPA INTERRATER AGREEMENT

|         | Interrater Agreement | Interpretation         |
|---------|----------------------|------------------------|
| Group 1 | 0.12                 | Slight agreement       |
| Group 2 | 0.91                 | Near perfect agreement |
| Group 3 | 0.67                 | Substantial agreement  |

TABLE II
DISTRIBUTION OF PRIVACY-RELATED QAS

|         | Both Annotators | One Annotator | Total |
|---------|-----------------|---------------|-------|
| Group 1 | 6               | 27            | 33    |
| Group 2 | 33              | 4             | 37    |
| Group 3 | 15              | 10            | 25    |
| Total   | 54              | 41            | 95    |

TABLE III
DISTRIBUTION OF PRIVACY-RELATED QAS - (SOLOVE [37])

| Privacy Concepts         | Our Dataset Distribution |
|--------------------------|--------------------------|
| Surveillance             | 20.65%                   |
| Interrogation            | 1.09%                    |
| Aggregation              | 30.43%                   |
| Insecurity               | 16.30%                   |
| Identification           | 27.17%                   |
| Secondary Use            | 3.26%                    |
| Exclusion                | 1.09%                    |
| Breach of Confidentiality| 4.35%                    |
| Disclosure               | 26.09%                   |
| Exposure                 | 10.87%                   |
| Increased Accessibility  | 1.09%                    |
| Blackmail                | 0%                       |
| Appropriation            | 1.09%                    |
| Distortion               | 0%                       |
| Intrusion                | 13.04%                   |
| Decisional Inference     | 0%                       |

## IV. RESULTS

In this section, we first discuss the results of our annotation process and our findings regarding the privacy-related questions asked on Stack Overflow and then provide the results related to the comparative analysis with ChatGPT. Our datasets and the results are publicly accessible here.[4]

### A. Classification Of Stack Overflow QAs

As mentioned in Section III, after Round_1 and revising the methodology, we divided 270 QAs into three groups. Each pair of annotators annotated 90 QAs, in total. After completing the four rounds, the total number of annotated privacy-related questions by both annotators was 54 and another 41 QAs were annotated as privacy by only one annotator. For each pair of annotators, we calculated the Cohen-Kappa interrater agreement prior to the Discussion_2. Table I shows the agreements. The agreements for Groups 2 and 3 are substantial to near-perfect while the agreement for Group 1 is counted as slight agreement. We analyzed the reason for the lower value for Group 1. We noticed that it is due to the imbalance level of expertise between the two annotators in this group (where one annotator has the highest level of privacy expertise whereas the other one is the least experienced among all six annotators). On the other hand, the other two groups were more balanced in terms of their privacy expertise. We mitigated this problem through the second round of discussion (see Section III.A for more details). After Discussion_2, the annotators decided that 92 QAs are privacy-related which is about 34.1% of the total annotated QAs. Tables III and IV show the distribution and the percentage of annotated privacy-related QAs.

We also asked the annotators to collect the accepted and/or the highest upvote answer. In total, 48/92 QAs have both accepted and highest upvote answers while the rest of the QAs has either one of them. Among those 48 that have both types of answers, for 42 of them, the accepted answer is the same as the highest upvote answer.

The annotators used the privacy concepts and the list of hypotheses from Harkous et al. [27] to determine and label privacy-related QAs. Note that, a QA may have more than one privacy concept. Hence, we end up with multi-label QAs in our dataset. Tables III and IV show the distribution of the labels in our dataset for Solove's and Wang and Kobsa's taxonomies. As shown in these two tables, the three most common categories are related to "choice/consent" (∼37% (34/92)), "aggregation" (∼30% (28/92)), and "identification" (∼27% (25/92)) of the QAs. As mentioned earlier, other related work identified that the topics of confidentiality, hide, inform and control were the most common advice given and that Apple and Google privacy labels had an impact on developers' privacy behaviors. Our results are aligned with these findings and trends. The next three highest common categories are "disclosure" (24/92), "onward transfer" (23/92), and "purpose specification" (21/92) which each constitute about 23-27% of the total QAs. These three categories are related to third-party sharing/disclosure which indicates that analysts and developers face challenges regarding when, for what purpose, and how to share data with third parties.

Furthermore, we evaluated the correlation between the labels. We noticed that "collection limitation" is always labeled together with "choice/consent" as well as "insecurity". ∼63% of questions annotated with "disclosure" are also annotated with "choice/consent" while in 25% of cases, it is also annotated together with "onward transfer". For half of the instances that are labeled with "aggregation", "onward transfer" is the second label. "Choice/consent" and "purpose specification" are labeled together in half of the instances, as well.

Concepts in Solove's privacy taxonomy are cited 142 times while the concepts in Wang and Kobsa's are cited 99 times. The pattern indicates that concepts in Solove's privacy taxonomy may be slightly more important to developers in general.

---
[4] https://tinyurl.com/so-gpt

TABLE IV
DISTRIBUTION OF PRIVACY-RELATED QAS - (WANG AND KOBSA [38])

| Privacy Concepts | Our Dataset Distribution |
|---|---|
| Notice/Awareness | 11.96% |
| Data Minimization | 4.35% |
| Purpose Specification | 22.83% |
| Collection Limitation | 3.26% |
| Use Limitation | 3.26% |
| Onward Transfer | 26.08% |
| Choice/Consent | 36.96% |

TABLE V
FLEISS-KAPPA INTERRATER AGREEMENT

|  | Interrater Agreement | Interpretation |
|---|---|---|
| Group 1 | 0.27 | Fair agreement |
| Group 2 | 0.27 | Fair agreement |
| Group 3 | 0.26 | Fair agreement |
| Group 4 | 0.41 | Moderate agreement |

*B. SOA/GPT-A Comparison*

After creating the datasets of SOA and GPT-A (as described in Section III), we quantitatively compared the responses between SOA and GPT-A. We computed the average word count between the two answers: the mean word count for SOA is 92.91 and for GPT-A is 288.81. This result suggests that ChatGPT generates more verbose and longer responses than the answers written by users on Stack Overflow.

Next, we divided the pair of SOA/GPT-A into four groups and distributed them among five out of the six original annotators as explained in Section III.B and ensured that each entry was seen by three annotators.

In the first step, the annotators compared the SOA with GPT-A to decide if the answers agreed with each other (Yes) or not (No). We calculated the Fleiss-Kappa interrater agreement for each group which indicates a fair to moderate agreement (see Table V). This agreement is only for the cases in which all three annotators agree on "Yes" or "No". However, in our work, we considered the answers with at least two "Yes" as an "agreed" answer. Our result shows that out of the 82 questions, 46 (56.1%) agree with each other (i.e., at least two annotators counted them as "agreed"). Among these 46 cases, for 21 of them, all three annotators "agreed". For the rest of the questions (i.e., 43.9%), either one or no annotators "agreed". The result shows that ChatGPT can generate acceptable responses in slightly more than half of the cases.

Next, each question labeled as "agreed" by at least two annotators was evaluated to determine whether the answers are an exact match (EM) (i.e., they propose the same solution) or a similar match (SM) (i.e., the solutions could be alternative or complementary). If two out of three annotators state that a pair of answers is an EM or SM, we label them as such. In cases where only two annotators label an answer as "agreed", the result for determining the EM or SM may be inconclusive (Inc). Out of the 46 questions with "agreed" answers, 20 (43.5%) of them are declared to be EM, and 23 (50%) of them are SM. 3 answer pairs were inconclusive.

Lastly, each question that did not have an "agreed" label was checked to determine if the SOA or GPT-A is more accurate. We again considered the majority of the votes (i.e., 2/3 of annotators) for the analysis. Since a small subset of questions was annotated by only two annotators as "disagreed", we ended up with a few inconclusive cases. Within the 36 questions with the "disagreed" label, the SOA is considered more accurate in 15 (41.7%) of them while the GPT-A is considered a more accurate answer for 12 (33.3%). 9 (25%) of the questions are marked inconclusive. Figure 5 shows the summary of the result.

|  | Agreed ||| Disagreed ||| Total SOA Acc | Total GPT-A Acc |
|---|---|---|---|---|---|---|---|---|
|  | EM | SM | Inc | SOA - Acc | GPT-A - Acc | Inc |  |  |
| Results | 20 | 23 | 3 | 15 | 12 | 9 | 61 (74%) | 58 (71%) |

Fig. 5. Comparison between SO and ChatGPT Answers

In total, SO responses are considered equal or more accurate for ~74% of the QAs, while for ChatGPT this number is ~71%. These results indicate that even though ChatGPT does not outperform SO yet, it could be used as an alternative solution in expert-required questions such as privacy. However, since privacy is a sensitive topic and non-compliance can result in fines and breaches, analysts and developers should not fully rely on these forums and tools for their questions.

V. DISCUSSIONS, CONCLUSION, AND FUTURE WORK

In this paper, we described a preliminary approach to extract and classify privacy-related questions from Stack Overflow (SO) and evaluated how ChatGPT compares with SO in assisting developers with privacy-related questions. We first extracted 932 SO entries posted between 2016 and 2023 and then randomly selected 270 pairs of QAs from the list. We then conducted four rounds of annotation process and two discussions by leveraging a manual NLI-based approach to classify them based on privacy concepts from the hypotheses list. Our annotation resulted in a multi-label dataset of 92 privacy-related QAs which we used to evaluate against ChatGPT-generated answers. Our initial results show that developers are more concerned with notice/consent, confidentiality, and third-party sharing and transfer. Furthermore, we found that, when given the same prompt, even though Stack Overflow slightly outperforms ChatGPT in providing more accurate responses, the results generated by ChatGPT could serve as a viable alternative for developers. Although our work provides a preliminary step toward a comprehensive analysis, there are some limitations that we plan to address in the future.

First, our annotation process is currently a manual process which is not only labor intensive but it may lead to inaccuracy and missing information. We also rely on *privacy* keyword for the initial selection of the data. Research has shown that relying on keyword searches and manual annotation may lead to missing a large number of privacy content (e.g., due to the use of different terminologies) and may result in overfitting the models on the presence or absence of privacy keywords in case of automation. We propose to use transformer-based NLI approaches to extract a larger pool of data, similar to [27], and then use text summarization [40] to identify the topics.

Second, the list of hypotheses used for our annotation process is tailored toward users. While privacy taxonomies are generic and agnostic toward users or developers, the hypothesis may still lead to some inaccuracy. We mitigated this limitation by using privacy experts as our annotators. However, in the future, we will modify the hypotheses list for fine-tuning and training our transformer-based model.

Third, we manually compared the result generated by ChatGPT with SO. Although using privacy experts for such comparison is helpful, with a larger dataset, this process can be time-consuming and error-prone. We plan to automate the process by leveraging similarity techniques, especially in cases where we have both text and code [41].

Fourth, we used the free version of ChatGPT which may result in inaccurate responses, in some cases. While most developers in small companies have access to the free version, comparing the results with GPT-4 version and/or other tools such as Bard could be insightful as well.

Lastly, we used the questions from SO *as-is* in ChatGPT to conduct a direct comparison. Developers may refine the prompts a few times to improve the accuracy of the results. In the future, for the cases in that ChatGPT underperformed, we will examine how many iterations are needed to improve the accuracy of the responses and provide some insights.

ACKNOWLEDGMENT

We would like to thank Sam Morse for their contribution to this work.REFERENCES

[1] A. Ekambaranathan, J. Zhao, and M. Van Kleek, ""money makes the world go around": Identifying barriers to better privacy in children's apps from developers' perspectives," in *Proc. of the 2021 CHI Conf. on Human Factors in Computing Systems*, 2021, pp. 1–15.
[2] S. Spiekermann, J. Korunovska, and M. Langheinrich, "Inside the organization: Why privacy and security engineering is a challenge for engineers," *Proc. of the IEEE*, vol. 107, no. 3, pp. 600–615, 2018.
[3] S. Spiekermann-Hoff, J. Korunovska, and M. Langheinrich, "Understanding engineers' drivers and impediments for ethical system development: The case of privacy and security engineering," 2018.
[4] A. Dalela, S. Giallorenzo, O. Kulyk, J. Mauro, and E. Paja, "A mixed-method study on security and privacy practices in danish companies," *ArXiv*, vol. abs/2104.04030, 2021.
[5] M. Green and M. Smith, "Developers are not the enemy!: The need for usable security apis," *IEEE S&P*, vol. 14, pp. 40–46, 2016.
[6] M. Tahaei and et al., ""i don't know too much about it": On the security mindsets of computer science students," in *Socio-Technical Aspects in Security and Trust*. Springer, 2021, pp. 27–46.
[7] N. Alomar, S. Egelman, and J. L. Fischer, "Developers say the darnedest things: Privacy compliance processes followed by developers of child-directed apps," *Proc. on PETS*, no. 4, 2022.
[8] K. Bednar and et al., "Engineering privacy by design: Are engineers ready to live up to the challenge?" *The Information Society*, vol. 35, no. 3, pp. 122–142, 2019.
[9] I. Hadar and et al., "Privacy by designers: Software developers' privacy mindset," *J. of Empirical Software Engineering*, vol. 23, no. 1, p. 259–289, 2018.
[10] Government of California, "California Consumer Privacy Act (CCPA)," https://oag.ca.gov/privacy/ccpa, accessed 2023).
[11] The EU, "The General Data Protection Regulation," http://www.eugdpr.org/, accessed, 2023.
[12] "Stack Overflow," https://stackoverflow.com/, (accessed 2023).
[13] "Reddit," https://www.reddit.com/r/programming/, (accessed 2023).
[14] N. Nascimento, P. Alencar, and D. Cowan, "Comparing software developers with chatgpt: An empirical investigation," 2023.
[15] H. Tian and et al., "Is chatgpt the ultimate programming assistant – how far is it?" 2023.
[16] M. Tahaei and et al., "Understanding privacy-related questions on stack overflow," in *Proc. of the 2020 CHI Conf.*, 2020, p. 1–14.
[17] M. Tahaei, T. Li, and K. Vaniea, "Understanding privacy-related advice on stack overflow," *PoPET*, vol. 2022, no. 2, pp. 114–131, 2022.
[18] T. Li, E. Louie, L. Dabbish, and J. Hong, "How developers talk about personal data and what it means for user privacy: A case study of a developer forum on reddit," *Proc. ACM Hum.-Comp. Inter.*, vol. 4, 2021.
[19] J. Parsons, M. Schrider, O. Ogunlela, and S. Ghanavati, "Understanding developers privacy concerns through reddit thread analysis," 2023.
[20] J.-H. Hoepman, "Privacy design strategies (the little blue book)," 2018.
[21] A. Radford, K. Narasimhan, T. Salimans, I. Sutskever *et al.*, "Improving language understanding by generative pre-training," 2018.
[22] T. Brown and et al., "Language models are few-shot learners," *Advances in neural information processing systems*, vol. 33, pp. 1877–1901, 2020.
[23] A. Fantechi, S. Gnesi, and L. Semini, "Rule-based nlp vs chatgpt in ambiguity detection, a preliminary study," 2023.
[24] J. White and et al., "Chatgpt prompt patterns for improving code quality, refactoring, requirements elicitation, and software design," *arXiv preprint arXiv:2303.07839*, 2023.
[25] D. Sobania and et al., "An analysis of the automatic bug fixing performance of chatgpt," *arXiv preprint arXiv:2301.08653*, 2023.
[26] S. R. Bowman and et al., "A large annotated corpus for learning natural language inference," *arXiv preprint arXiv:1508.05326*, 2015.
[27] H. Harkous, S. Peddinti, R. Khandelwal, A. Srivastava, and N. Taft, "Hark: A deep learning system for navigating privacy feedback at scale," in *IEEE Symp. on Sec. & Privacy*, 2022.
[28] J. White and et al., "A prompt pattern catalog to enhance prompt engineering with chatgpt," *arXiv preprint arXiv:2302.11382*, 2023.
[29] N. Proferes and et al., "Studying reddit: A systematic overview of disciplines, approaches, methods, and ethics," *Social Media+ Society*, vol. 7, no. 2, 2021.
[30] D. Greene and K. Shilton, "Platform privacies: Governance, collaboration, and the different meanings of "privacy" in ios and android development," *New Media & Society*, 2018.
[31] T. Iqbal, M. Khan, K. Taveter, and N. Seyff, "Mining reddit as a new source for software requirements," in *IEEE 29th Int. Requirements Engineering Conf.*, 2021, pp. 128–138.
[32] M. Tahaei, J. Bernd, and A. Rashid, "Privacy, permissions, and the health app ecosystem: A stack overflow exploration," in *The European Symp. on Usable Security*, 2022, p. 117–130.
[33] A. Felfernig and et al., Eds., *SPLC '22: 26th ACM International Systems and Software Product Line Conference, Graz, Austria, September 12 - 16, 2022, Volume A*. ACM, 2022.
[34] N. Nascimento, P. Alencar, C. Lucena, and D. Cowan, "Toward human-in-the-loop collaboration between software engineers and machine learning algorithms," 12 2018, pp. 3534–3540.
[35] C. Raffel and et al., "Exploring the limits of transfer learning with a unified text-to-text transformer," *Journal of Machine Learning Research*, vol. 21, no. 140, pp. 1–67, 2020.
[36] B. MacCartney and C. D. Manning, "An extended model of natural logic," in *Proceedings of the Eighth International Conference on Computational Semantics*, ser. IWCS-8 '09. ACM, 2009, p. 140–156.
[37] D. J. Solove, "A taxonomy of privacy," *University of Pennsylvania Law Review*, vol. 154, no. 3, pp. 477–564, 2006.
[38] Y. Wang and A. Kobsa, "Privacy-enhancing technologies," in *Handbook of research on social and organizational liabilities in information security*. IGI Global, 2009, pp. 203–227.
[39] A. Williams, N. Nangia, and S. Bowman, "A broad-coverage challenge corpus for sentence understanding through inference," in *Proc. of the Conf. of the North American Chapter of the ACL: HLT, V. 1*, 2018.
[40] L. Liu and et al., "Generative adversarial network for abstractive text summarization," 2017. [Online]. Available: https://arxiv.org/abs/1711.09357
[41] J. Akram and et al., "Droidcc: A scalable clone detection approach for android applications to detect similarity at source code level," in *2018 IEEE 42nd Annual COMPSAC*, vol. 01, 2018, pp. 100–105.